\documentclass{emulateapj}
\lefthead{GAUDI} \righthead{SIZE DISTRIBUTION OF CLOSE-IN PLANETS}

\def\eq#1{equation (\ref{#1})}

\def\mj{M_{J}}
\def\rj{R_{J}}

\def\sn{{\rm S/N}}

\def\ntot{N_{\rm tot}}

\def\teq{T_{\rm eq}}
\def\rpmi{\langle R_{p}\rangle}
\def\sigi{\sigma}
\def\rpmo{\langle R_{p}\rangle_o}
\def\sigo{\sigma_{o}}
\def\rp{R_p}
\def\sn{{\rm S/N}}

\begin{document}
\title{On the Size Distribution of Close-In Extrasolar Giant Planets}
\author{B.\ Scott Gaudi}

\affil{Harvard-Smithsonian Center for Astrophysics, 60 Garden St., Cambridge, MA 02138}
\email{sgaudi@cfa.harvard.edu}

\begin{abstract}
The precisions of extrasolar planet radius measurements are reaching
the point at which meaningful and discriminatory comparisons with
theoretical predictions can be made.  However, care must be taken to
account for selection effects in the transit surveys that detect the
transiting planets for which radius measurements are possible.  Here I
identify one such selection effect, such that the number of planets
with radius $\rp$ detected in a signal-to-noise ratio limited transit survey
is $\propto \rp^\alpha$, with $\alpha\sim 4-6$.  In the presence of a
dispersion $\sigi$ in the intrinsic distribution of planet radii, this
selection effect translates to bias $b$ in the radii of observed
planets.  Detected planets are, on average, larger by a fractional
amount $b\sim \alpha (\sigi/\rpmi)^2$ relative to the mean radius
$\rpmi$ of the underlying distribution.  I argue that the intrinsic
dispersion in planetary radii is likely to be in the range $\sigi =
(0.05-0.13)\rj$, where the lower bound is that expected theoretically
solely from the variance in stellar insolation, and the upper bound is
the 95\% c.l.\ upper limit from the scatter in observed radii.
Assuming an arbitrary but plausible value of $\sigi/\rpmi \sim 10\%$,
and thus $b\sim 6\%$, I infer a mean intrinsic radius of close-in
massive extrasolar planets of $\rpmi=(1.03 \pm 0.03)\rj$.  This value
reinforces the case for HD209458b having an anomalously large radius,
and may be inconsistent with coreless models of irradiated giant
planets.
\end{abstract}

\section{Introduction}\label{sec:intro}

After many years of effort, transit searches for extrasolar planets
have recently come to fruition, with the detection of six extrasolar
planets via the transit method.  Most of these planets were discovered
in deep Galactic field surveys by the OGLE collaboration
\citep{udalski02a,udalski02b,udalski02c,udalski03}, and subsequently
confirmed via radial velocity follow-up
\citep{konacki03a,konacki03b,konacki04,konacki05,
bouchy04,pont04,moutou04,pont05,bouchy05}.  One was discovered in a shallow,
wide-angle field survey by the TrES collaboration \citep{alonso04}.
Along with the transiting planet HD209458b
\citep{henry00,charbonneau00}, originally discovered via radial
velocity surveys \citep{mazeh00}, seven transiting planets are
currently known, with many more sure to follow.

The photometric light curve of a star with a transiting planet, when
combined with detailed spectroscopic follow-up including precise
radial velocities, yields the planet radius $\rp$, mass $M_p$, and
semimajor axis $a$, along with the mass $M_*$, radius $R_*$, and
effective temperature of the host star.  
Of the
routinely observable properties of transiting extrasolar planets, the
radius is the primary diagnostic for testing theoretical models.
Currently, the radii of the seven known
transiting planets are determined to precisions of $\Delta\rp/\rp\equiv\delta\rp\sim(5-10)\%$.
A number of different sources of uncertainty contribute to 
the total error, 
but ultimately the uncertainty in the radius of a transiting planet is limited
by the uncertainty in the primary mass, such that $\delta \rp \sim (1/3)\delta M_*$.
Stellar masses can be inferred with precisions of $\sim(10-20)\%$, and thus,
in the future, we can expect to be able to measure planetary radii
to $\sim(3-7)\%$. 

The steadily increasing number and precision of planet radius
measurements, along with improvements in the theoretical models of
irradiated giant planets, has led many authors to present detailed
comparisons between measured radii and model predictions.  Other than the mysterious case of
HD209458b, whose anomalously large radius continues to elude a
satisfying explanation \citep{gs02,burrows03,baraffe03,deming05,winn05}, these
comparisons have generally resulted in agreement between
theory and observations 
\citep{bll03,burrows04,chabrier04,guillot05,laughlin05}.  Although
detailed inferences about the ensemble properties of giant close-in
planets are not yet possible, the expected precisions should be sufficient to
distinguish between, e.g., models with and without a large solid core.

Before celebrating the agreement between the theoretical
predictions and observational constraints on extrasolar planetary
radii, we should be certain that all relevant biases 
in the observations and/or theoretical predictions
have been identified.  \citet{burrows03} identified one such bias, such
that the planet radius as inferred from a transit light curve is $\sim(3-10)\%$ 
larger than the photospheric radius predicted by models, due
to the fact that rays from the primary passing perpendicular to the planet 
radius vector suffer a longer path length through the atmosphere.   
The purpose of this {\it Letter} is to point out
that there exists an additional selection effect on
the radii of planets detected in signal-to-noise ratio ($\sn$) limited field transit
surveys. This selection effect is such that the number of detected
planets is $\propto R_p^\alpha$, with $\alpha \sim 4-6$.  In the
presence of intrinsic scatter in the distribution of planetary radii,
this translates directly into a bias in the mean radius of
detected planets relative to the intrinsic population.  This bias can
affect comparisons of theoretically-predicted radii with the ensemble
distribution of observed planet radii.  It will also
affect the interpretations of individual systems 
if there exist unaccounted-for dependences
of the radius on unobservable parameters of the system, such as the migration
timescale or an indeterminate amount of internal heating. Note that
the bias identified here operates in the {\it same} sense as the
\citet{burrows03} `transit radius' effect.  Therefore, these two biases
can easily combine to yield $\sim 10\%$ differences from the usual
naive comparisons.  These biases are easy to understand.  
However, they must be
acknowledged and considered when drawing conclusions about the
agreement (or lack thereof) between theory and observations.

\section{Selection Effects in Field Transit Surveys}\label{sec:selection}

Field surveys for transiting planets are subject to a number of
selection effects, which can lead to observed distributions of
planetary parameters that are biased with respect to the underlying
intrinsic distributions.  These selection effects have been discussed
by \citet{gaudi05} and \citet{pont05}, although these studies were primarily
concerned with biases in the periods of detected extrasolar planets
and did not consider the bias with respect to planetary radius in any
detail. \citet{pgd03} and \citet{gaudi05} present simple scaling relations for
the number of planets detected in a $\sn$-limited field
transit survey, as a function of the parameters of the star and
planet.  I review the basic steps
here, but refer the reader to these papers for a more detailed
derivation of these relations.

The total $\sn$ of the photometric detection
of a  planetary transit is,
\begin{equation}
{\rm \frac{S}{N}}= N_{tr}^{1/2} \left(\frac{\delta}{\sigma_{ph}}\right),
\label{eqn:sntrans}
\end{equation}
where $N_{tr}=(R_*/\pi a)\ntot$ is the total number of points in
transit, $\ntot$ is the total number of data points in the light
curve, $\delta=(\rp/R_*)^2$ is the depth of the transit, and $\sigma_{ph}$
is the single-measurement relative photometric precision.  In terms of
fundamental parameters of the planet and star, the $\sn$ scales as
\begin{equation}
{\rm \frac{S}{N}} \propto R_*^{-3/2} \teq^2 R_p^2 a^{1/2} d^{-1},
\label{eqn:snfund}
\end{equation}
where $\teq$ is the equilibrium temperature of the planet,
and I have assumed that $\sigma_{ph}$ is limited by Poisson noise, such that $\sigma_{ph}\propto
L_*^{-1/2} d$, where $L_*$
and $d$ are the luminosity and distance of the
host star, respectively.  

I note that the various factors in \eq{eqn:snfund} are
unlikely to be independent.  For example, it may be that planet properties are correlated with host
star mass and thus $R_*$.  In addition,
models of irradiated planets
predict that their radii will likely depend on the amount of stellar insolation
absorbed by the planet.  Thus $\rp$ is expected to be correlated
with $\teq$, although precise predictions for this correlation are
hampered by an incomplete understanding of the physical processes
involved with the absorption and redistribution of the energy from
the incident stellar radiation.  There is also observational evidence that
the mass of close-in
giant planets is correlated with $a$, such that planets closer to
their parent star are more massive \citep{gaudi05,mazeh05}.  All
else equal, this would lead to a correlation of $\rp$ with
$a$.  \citet{laughlin05} argue that such a correlation between
$\rp$ and $a$ largely explains the apparent discrepancy between the
period distributions inferred from radial velocity surveys and transit
surveys.  This is unlikely to be correct, first because the
evidence for a correlation of $\rp$ with $a$ is weak, and second
because, even for the strength of the correlation quoted by
\citet{laughlin05}, the magnitude of resulting selection effect would
be less than half that due to the inevitable selection effect
arising from the direct dependence on $a$ \citep{gaudi05}.

Although these correlations are interesting topics
for future study, their contribution to the selection effects 
are unlikely to dominate over the direct $R_p^2$ term  in \eq{eqn:snfund}.  
I will therefore ignore the other terms for the remainder of the discussion,
and assume $\sn \propto R_p^2 d^{-1}$.

At a limiting
$(\sn)_{\rm min}$, the maximum stellar distance out to which a transiting planet can be detected
is therefore $d_{\rm max} \propto \rp^2$.
The number of planets detected
is  proportional to the total volume over which a planet gives rise to a
transit with $\sn\ge(\sn)_{\rm min}$.
Assuming a constant volume density $n$ of stars and no dust, the number of planets that
can be detected is therefore $\propto d_{\rm max}^3 \propto \rp^6$.   
If the intrinsic distribution of planetary radii is $f_i(\rp)\equiv dn/d\rp$,
the observed distribution of planetary radii $f_o(\rp)$ will be
\begin{equation}
f_o(\rp) \propto f_i(\rp)\rp^\alpha,
\label{eqn:fobs}
\end{equation}
with $\alpha =6$.  I leave $\alpha$ as a free parameter to allow for 
departures from the ideal case, as discussed below.

There are a number of assumptions that enter into the derivation of
\eq{eqn:fobs}, some of which I have explicitly stated, and others of
which are reviewed in \citet{gaudi05}.  The most relevant for the
present discussion are the assumptions of a constant volume density of
stars, no dust, Poisson-noise limited photometry, and a
$\sn$-limited survey.  All of these are violated to some
extent in the actual field surveys that have detected transiting
extrasolar planets.  S.\ Dorsher, A.\ Gould, \& B.\ S.\ Gaudi (in preparation)
calculate the expected scaling
of the number of detected planets with $\rp$ for the OGLE surveys,
using a realistic Galactic model of the source star and dust
distribution, including the joint distribution of host star
luminosities and radii, and considering the actual error properties of
the OGLE photometry.  They find that the number of detected planets is
a high power of $\rp$, but with a somewhat smaller exponent than $\alpha=6$.
The range is $\alpha \sim 4-6$, with lower indices expected for
planets with smaller $a$.  The properties of the TrES survey have not
been described in detail, so I will assume for simplicity that they
are similar to the OGLE survey.

\section{The Radius Bias}\label{sec:bias}

It is clear from \eq{eqn:fobs} that due to the $\rp^\alpha$ selection effect
in $\sn$-limited transit surveys, the observed distribution of
planet radii will be a biased subset of the underlying intrinsic distribution. 
For example, the mean of the observed distribution will generally be larger than the 
mean of the underlying distribution.  As a result, if this
bias is not taken into account, incorrect inferences about the population of planets
as a whole could be drawn from the properties of observed planets.

\begin{figure}
\epsscale{1.0} 
\plotone{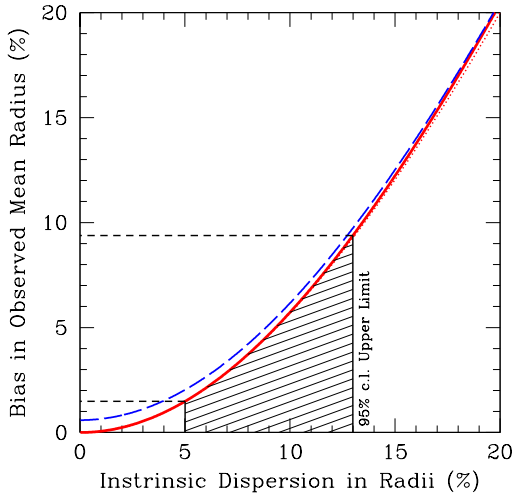}
\caption{\label{fig:one}
Bias vs. intrinsic dispersion in radii.  The solid curve shows the
fractional difference between the observed and intrinsic mean radius
(the bias) versus the intrinsic dispersion in radii, assuming the
intrinsic distribution is a Gaussian.  The dotted curve shows a simple
approximation derived from \eq{eqn:rmpoapp}.  
The long-dashed curve shows the bias assuming 
there exists an additional population of planets whose radii are inflated by $10\%$,
and constitute $\sim 10\%$ of the total population of planets.  
The shaded
region shows the likely range of intrinsic dispersion in radii.
}
\end{figure}

The magnitude of the bias will clearly depend on the intrinsic
distribution of planets.  To provide a quantitative estimate of the
bias, I adopt a Gaussian form for the intrinsic distribution $f_i$ 
of planet radii,
with mean $\rpmi$ and standard deviation $\sigi$.  The mean $\rpmo$
and standard deviation $\sigo$ of the observed planet distribution
$f_o$ can then be calculated in the usual way.  Although exact
analytic expressions for $\rpmo$ and $\sigo$ can be found, they are
cumbersome and not particularly illuminating.  The bias, defined here
as the fractional difference between the mean of the observed distribution
relative to the mean of the intrinsic distribution, $b\equiv
(\rpmo-\rpmi)/\rpmi$, is shown in Figure \ref{fig:one}.  For
$\sigi/\rpmi\ll 1$, the observed distribution $f_o$ is approximately a
Gaussian.  
Therefore, the mean of $f_o$ can be approximated by its
maximum, which yields,
\begin{equation}
\rpmo \simeq \rpmi \left[\frac{1}{2} +
\frac{1}{2}\left(1+4\alpha\frac{\sigi^2}{\rpmi^2}\right)^{1/2}\right].
\label{eqn:rmpoapp}
\end{equation}
The resulting bias using this approximation is shown in Figure
\ref{fig:one}; it agrees well with the exact result.  For $\sigi/\rpmi \ll
(4\alpha)^{-1/2}$, 
\begin{equation}
b\simeq \alpha\left(\frac{\sigi}{\rpmi}\right)^2.
\label{eqn:bias}
\end{equation}
Thus, if the intrinsic distribution of planet radii has a standard
deviation of $10\%$, the mean of the observed planets will be larger
than that of the underlying population by $\sim 4-6\%$.  I note that the
$\rp^\alpha$ selection effect also tends to result in an 
observed dispersion that is smaller than the intrinsic dispersion.  
However, for $\sigi/\rpmi \la 25\%$, the intrinsic and observed dispersions
differ by $\la 10\%$ for $\alpha \le 6$. 

The magnitude of the bias is relatively insensitive to the form of the intrinsic
distribution $f_i(\rp)$.  As an example, consider the case where bulk
of planets have the form assumed in the above analysis, but there exists an
additional population of ``inflated'' planets.  The
inflated planets have the same dispersion as the other planets, but 
their mean radii are larger by $10\%$, and are ten times less common.
Such a population might be indicated by the existence of HD209458b,
which appears to have an anomalously large radius.  In this
case, the bias is very similar, as shown in Figure
\ref{fig:one}.

\begin{figure}
\epsscale{1.0} 
\plotone{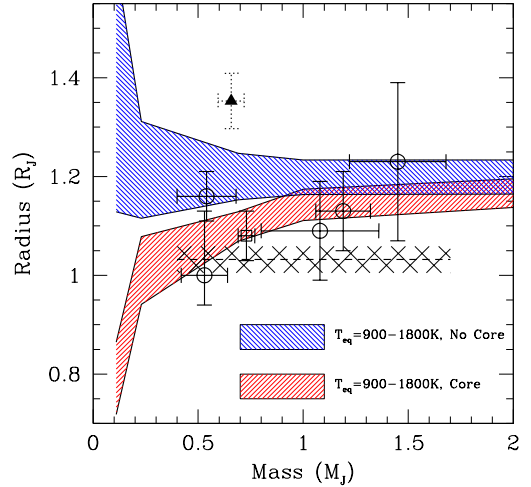}
\caption{\label{fig:two} 
Radius versus mass for the seven known transiting extrasolar planets.
The circles show planets confirmed from follow-up observations of
candidates from the OGLE deep field survey, while the square shows
TrES-1, which was detected from a wide-angle, shallow field survey.  The
triangle shows HD209458b, which was originally discovered using radial
velocities, and as such is not subject to the same biases as the
planets detected in field surveys.  The shaded regions show the
predictions of \citet{bll03} for radii of irradiated planets with equilibrium
temperatures in the range expected for the known transiting planets,
$\teq=900-1800~{\rm K}$. These radii have been augmented by 5\%, to 
roughly account
for the `transit radius' effect discussed in \citet{burrows03}.  The dashed 
line and hatched region shows the mean
radius and error of the intrinsic population of extrasolar planets,
as inferred from `debiasing' the data from the six planets
detected in field surveys, assuming an intrinsic dispersion in radii of $10\%$.}
\end{figure}

\section{Summary and Discussion}\label{sec:disc}

I have argued that there exists a bias in the radii of extrasolar
planets detected in field transit surveys.  This bias arises from the
fact that there exists a strong selection effect in these surveys, such
that the number of planets detected is $\propto \rp^\alpha$, with
$\alpha \sim 4-6$.  The magnitude of this bias depends on the form of
the intrinsic distribution of planet radii, and is $b \sim \alpha
(\sigi/\rpmi)^2$ for a Gaussian distribution with mean $\rpmi$ and
dispersion $\sigi$.  Although the assumption of a Gaussian distribution
of radii is reasonable, I stress that a bias exists for {\it any}
intrinsic distribution with a finite dispersion. 
 
One might be tempted to conclude that this bias only effects
inferences about the ensemble distribution of planetary radii, and
that point-to-point comparisons between calculations 
and measurements of the radii of individual systems would be
immune to bias.  This is not necessarily the case, however, because there may
exist unobservable or poorly constrained parameters that effect the
planetary radius, such as the migration timescale, age of the system,
or an indeterminate amount of internal heating.  These hidden
parameter dependences may give rise to a dispersion in planetary radii
at fixed values of the observable parameters.  Thus an observed planet
is likely to have a larger radius than might be expected based on models
incorporating typical values of these hidden parameters. 

Although the primary purpose of this paper is to simply point out the
existence of a radius bias, it is interesting to reconsider the comparison
between models and data in light of
this previously unaccounted-for effect. 
To do so, I use the
predictions of models of irradiated giant planets by \citet{bll03} and
\citet{laughlin05}.  Predictions of other models
are similar, although there are discrepancies at the $\sim 10\%$ level. 
I use these models simply because these authors
present their predictions in an easily accessible numerical form,
specifically as tables of predicted radii as a function of planet mass
$M_p$ and $\teq$ \citep{bll03}, and as predictions for specific
systems \citep{laughlin05}. 
These models span planet masses in the
range $M_p=(0.11-3)\mj$, equilibrium temperatures in the range
$\teq=(113-2000){\rm K}$, and consist of models with and without
solid cores of mass $20~M_\oplus$ for $M_p< \mj$ and $40~M_\oplus$ for
$M_p \ge \mj$.  I increase all model radii by $5\%$ to roughly account
for the `transit radius' effect discussed by \citet{burrows03}.  This
is somewhat larger than the magnitude expected for more massive
planets such as OGLE-TR-56b \citep{burrows04}, and about half that
expected for less massive planets like HD209458b \citep{burrows03}.  I
augment the specific predictions of \citet{laughlin05} to include the
recently-confirmed planet OGLE-TR-10, using the stellar and planet
parameters from \citet{konacki05}, but accounting for the improved
photometry of \citet{holman05}.  Linearly
interpolating from Table 1 of \citet{bll03}, the predictions for the
radius of this planet ($\teq=1317~{\rm K}$) are $\rp=1.04~\rj$ and
$\rp=1.18~\rj$ for models with and without a core, as compared to the
(preliminary) measured radius of $\rp=(1.16 \pm 0.05)\rj$.

Figure \ref{fig:two} shows the radii of observed extrasolar planets,
together with the predictions of \citet{bll03} for the range of $\teq$
spanned by these systems, $\teq \simeq (900-1800){\rm K}$.  As has
been noted by numerous authors, the radius of HD209458b is anomalously
large.  It is important to emphasize that this object is {\it not}
subject to the bias discussed here, as it was originally discovered
via radial velocity measurements.  The average {\it observed} radius
of the remaining six planets is $1.10\rj$, with an error in the mean
of $0.03\rj$ and an RMS of $0.06\rj$.  The intrinsic dispersion is
consistent with zero, however the true dispersion is likely
considerably larger.  Just from the variance in $\teq$, the models of
\citet{bll03} predict a dispersion of $(3-5)\%$.  Any additional
variance in the properties of the systems that affect the planet
radius (i.e.\ in the age, metallicity, or core mass) would give rise
to a larger dispersion.  The 95\% c.l.\ upper limit on the
intrinsic dispersion is $0.13~\rj$, respectively.  Thus
the intrinsic dispersion is likely in the range $0.05-0.13~\rj$.

Without detailed knowledge of the intrinsic scatter in the radii of
giant planets, it is difficult to assess the effect of the bias on the
current dataset with any confidence.  I will therefore proceed rather
speculatively.  For definiteness, I will adopt $\sigi=0.1~\rj$.  
Although arbitrary,
this value seems plausible given expectations about the variance in
planet properties and histories.  Assuming $\alpha=6$, this gives a
bias of $b\simeq 6\%$.  Using this to `debias' the mean radius of the
observed planets gives an estimate of the mean intrinsic radius of
close-in massive extrasolar planets of $\rpmi = (1.03 \pm 0.03)\rj$.
In contrast, the models of \citet{bll03} and \citet{laughlin05}
predict $\rpmi=(1.16\pm 0.01)\rj$ for coreless planets with the
observed masses and effective temperatures.  
Provided that the models of \citet{bll03} span the full range
of intrinsic planet properties, then this inferred value of the mean
intrinsic radius implies that the majority of planets have a massive
core.  This also strengthens the case for HD209458b having a
anomalously large radius.

Although these results are tantalizing, 
I stress that the comparison between the measured radii and
theoretical predictions presented here is only preliminary.  A more
detailed study should be performed using a full suite of theoretical
predictions, and a more careful consideration of the observational and
theoretical biases.

\acknowledgments 
This work was supported by a Menzel Fellowship from the Harvard College
Observatory. I would like to thank Matt Holman, Dimitar Sasselov, and Josh Winn
for valuable comments on the manuscript and general
encouragement, and the anonymous referee
for a prompt and helpful report.

\end{document}